\documentstyle[prd,aps]{revtex}
\begin{document}
\draft
%%%%%%%%%%%%%%%%%%%%%%%%%%%%%%%%%%%%%%%%%%%%%%%%%%%%%%%%%%%%%%%%%%%%%%
%
%  Uncomment following two lines and one below for 2 column format.
%
\twocolumn[\hsize\textwidth\columnwidth\hsize\csname
@twocolumnfalse\endcsname
%%%%%%%%%%%%%%%%%%%%%%%%%%%%%%%%%%%%%%%%%%%%%%%%%%%%%%%%%%%%%%%%%%%%%%

\date{July 1997}
\title{\textbf{Magnetic Fields from Phase Transitions}}
\author{Mark Hindmarsh\cite{mhaddress}
}
\address{
Centre for Theoretical Physics\\
University of Sussex\\
Falmer\\
Brighton BN1 9QJ\\
U.K.}
\author{Allen Everett\cite{aeaddress}}
\address{Tufts Institute of Cosmology\\
Department of Physics and Astronomy\\
Tufts University\\
Medford MA 02155\\
U.S.A.}
\preprint{SUSX-TH-97-011, astro-ph/9708004}

\maketitle
%\vspace{-11cm}
%\begin{flushright}
%SUSX-TH-97-011\\
%astro-ph/9708004
%\end{flushright}
%\vfill

\begin{abstract}
The generation of primordial magnetic fields from
cosmological phase transitions is discussed, paying particular attention to the
electroweak transition and to the 
various definitions of the `average' field that have been put forward.
It is emphasised that only the volume average has dynamical significance
as a seed for galactic dynamos. 
On rather general grounds of causality and energy conservation, it
is shown that, in the absence 
of MHD effects that transfer power in the magnetic field from small to 
large scales, processes occurring at the electroweak transition 
cannot generate fields stronger than $10^{-20}$ Gauss on 
a scale of 0.5 Mpc.
However, it is implausible that this upper bound 
could ever be reached, as it would require all the energy in
the Universe to be turned into a magnetic field coherent
at the horizon scale.  Non-linear MHD effects seem therefore 
to be necessary  
if the electroweak transition is to create a primordial seed field.
\end{abstract}

\pacs{PACS numbers: 98.80.Hw \hfill Preprint: SUSX-TH-97-011, astro-ph/9708004}

%  This is the other line to be uncommented for 2 column format
\vskip2pc]
%%%%%%%%%%%%%%%%%%%%%%%%%%%%%%%%%%%%%%%%%%%%%%%%%%%%%%%%%%%%%%%%%%%%%%

\section{Introduction}

The origin of galactic magnetic fields appears to lie in a primordial 
seed field, produced in the very early Universe. 
The primordial magnetic flux would have been frozen into the
highly conductive plasma on scales sufficiently large that the diffusion of
the magnetic field can be neglected, keeping constant the magnetic flux through a closed
curve fixed in the plasma. This field would have 
been trapped and compressed in the collapsing protogalaxy, and then 
amplified to the observed intragalactic value, of order $10^{-6}$ G, by a
dynamo effect during galaxy formation \cite{MagF}.
It is believed that the galactic dynamo needs only 
a very small field, perhaps of
order $10^{-18}$ G to $10^{-21}$ G, to operate.  Given that a galaxy 
represents an overdensity of order $10^{-6}$, the primordial field 
as it would be measured today could be a factor $10^{4}$ smaller.

There seems to be no shortage of ways in which fundamental processes 
in the early Universe could have produced such a seed field.  
Most fall into two classes, according to whether 
the field is produced during an inflationary era or during a phase
transition, although a mechanism involving asymmetric neutrino
emission from rotating black holes manages to avoid this classification
\cite{VilLea81}.

Turner and Widrow \cite{TurWid88} estimated the production of 
large-scale magnetic fields in a set of inflationary models.  Ordinary
electromagnetism is conformally invariant, and therefore the initial
vacuum state remains the vacuum during inflation, and there can be no
particle production.  They therefore considered a number of models with
broken conformal invariance, but 
the only one which produced a large enough seed field
did not respect gauge invariance.  
Ratra \cite{Rat92} considered what is perhaps the most promising
inflationary model, where the inflaton couples to the electromagnetic
field in a dilaton-like manner.  Dolgov \cite{Dol93} pointed out that
this coupling was inevitably present due to the trace anomaly, although
one would require a huge imbalance in the numbers of bosonic and
fermionic degrees of freedom to generate the size of seed field that
seems to be required.

More recently, Mazzitelli and Spedalieri 
\cite{MazSpe95} have shown that there are (gauge-invariant) models 
with higher-derivative couplings which produce large enough seed fields,
although they involve derivative couplings of a rather high order. Still
with inflation, Gasperini et al. \cite{GasGioVen95} and Lemoine and 
Lemoine \cite{LemLem95} used the fact that theories based 
on the low-energy limit of string theory are not conformally invariant
in the electromagnetic sector because of the interaction with the
dilaton, and thus could in principle generate large-scale fields 
during inflation, given a big enough change in the value of the 
dilaton field. However, the two sets of authors differed over whether 
a sufficiently large change occurs in the string-inspired  
``pre-Big-Bang'' scenario of Gasperini and Veneziano \cite{GasVen93}. 

The first to suppose that a cosmological phase transition may generate a
magnetic field appears to have been Hogan \cite{Hog83}.  His 
idea was that an unspecified mechanism might operate 
during a first-order phase transition, which could generate a field 
with energy density of order the equipartition value, coherent on the
scale determined by the size of bubbles of the low temperature phase. 
Hogan concentrated on the QCD phase transition at 0.2 GeV. 
Since then, several specific mechanisms for generating 
a seed field at first order phase transitions 
have been put forward \cite{BayBodMcL96,SigOliJed97,AhoEnq97}.
The attractive possibility that the required
seed field could be produced in the electroweak phase transition,
even a continuous one, 
was first introduced by Vachaspati \cite{Vac91}.  
His estimate put the 
size of the field way below that necessary to supply the seed for 
galactic fields. However, subsequent 
estimates by another group \cite{EnqOle93} 
were several orders of magnitude larger, and they came to the opposite 
conclusion: 
that the electroweak phase transition {\em could} produce the required seed 
field.  There is therefore some confusion about the subject.

It is the aim in this paper to try and remove some of this confusion
concerning the generation of magnetic fields at the electroweak and 
other phase
transitions.  We begin by reprising Vachaspati's and Enqvist and Olesen's
arguments.  We discuss the definition of the magnetic
field tensor in the electroweak theory, which is not unique, and
the averaging procedures used by both sets of authors.  We find that
although Enqvist and Olsen perform the calculation of the line-averaged 
magnetic field correctly, the result is not the one required to estimate
the size of the seed field.  It is the {\em volume} average over galactic 
scales that is needed. 

It is further pointed out that, independently of the 
definition of the electromagnetic field, one would in any
case expect to find a thermal magnetic field at the transition,
with correlation length $(eT)^{-1}$ (the inverse Debye mass). 
Any primordial seed field must be clearly distinguished both 
in scale and amplitude from the thermal fluctuations, which 
have evolved into the microwave
background radiation, whose fields are negligible 
at galactic scales. It follows that any seed field must  
be generated by a non-equilibrium process. Energy
conservation and causality put strong limits on the strength 
of fields generated at phase transitions independently of
the details of the process. 

\section{Magnetic fields from phase transitions}

In a phase transition in the early Universe, a multicomponent Higgs scalar
field $\Phi $ (which may be fundamental or composite) acquires a vacuum
expectation value of modulus $\eta \simeq T_{\textrm{c}}$, where $T_{\textrm{c}}$
is the critical temperature at which the phase transition occurs. In the
case of the electroweak transition, let $W_\mu ^a$ ($a=1,2,3$) be the three
SU(2) vector boson fields and $Y_\mu $ the U(1) gauge boson, which have
field strengths $F_{\mu \nu }^a$ and $F_{\mu \nu }^0$ respectively. The
electromagnetic vector potential $A_\mu $ may then be defined by  
\begin{equation}
\label{eq1}A_\mu =-W_\mu ^a\widehat{\phi }^a\sin \theta _{\textrm{w}}+Y_\mu
\cos \theta _{\textrm{w}},
\end{equation}
where $\theta _{\textrm{w}}$ is the Weinberg angle and $\widehat{\phi }^a=\Phi
^{\dagger }\sigma ^a\Phi /|\Phi |^2$ is a unit SU(2) vector, which is
well-defined barring topological obstructions \cite{Nam77,HinJam94}. 

Using this unit isovector, we can also project out the linear combination of
the field strength tensors associated with the electromagnetic field:%
$$
F_{\mu \nu }^{\textrm{em}}=-F_{\mu \nu }^a\widehat{\phi }^a\sin \theta _{\textrm{%
w}}+F_{\mu \nu }^0\cos \theta _{\textrm{w}}. 
$$
It might seem natural that this is the field strength of electromagnetism:
however, there is another gauge invariant quantity with the correct symmetry
properties, namely
$$
D_{\mu \nu }=\epsilon ^{abc}\widehat{\phi }^aD_\mu \widehat{\phi }^bD_\nu 
\widehat{\phi }^c, 
$$
where $D_\mu \widehat{\phi }^a=(\partial _\mu +g\epsilon ^{abc}W_\mu ^b)%
\widehat{\phi }^c$. It is not immediately obvious how to combine the two
quantities to make an electromagnetic field strength tensor.
Vachaspati \cite{Vac91}
follows 't Hooft's convention \cite{tHo74}, which is to use%
\begin{equation}
{\cal F}_{\mu \nu }^{\textrm{em}}=F_{\mu \nu }^{\textrm{em}}+\frac{\sin \theta _{%
\textrm{w}}}gD_{\mu \nu }. 
\label{eFtH}
\end{equation}
This has the advantage that it obeys the Bianchi identity (which includes
the Maxwell equation $\nabla \cdot {\bf B}=0)$ almost everywhere, with the
exception of isolated points where the isovector field vanishes with
non-zero index. This, however, is not a particularly good criterion, as
we know that magnetic charge exists in non-Abelian theories. A 
further argument against this definition come from the fact 
that the energy density of the
electromagnetic field is no longer $\frac 12({\bf E}^2+{\bf B}^2)$. 
This serves to emphasise the point that the definition is more
a matter of taste than physics, and arguments which rest on a 
particular choice are on shaky ground.

In any case, let us reproduce Vachaspati's argument for the production of
magnetic fields by the electroweak phase transition. The minimum energy
state of the Universe corresponds to a spatially homogeneous vacuum in which 
$\Phi $ is covariantly constant, i.e. $D_\mu \Phi =0,$ with $D_\mu =\partial
_\mu -igW_\mu ^a\tau ^a-ig^{\prime }Y_\mu $. It also follows that $D_\mu 
\widehat{\phi }^a=0$. In this state there are no electromagnetic (or any
other) excitations. However, immediately after the phase transition there is
a finite correlation length $\xi $, which means that the Higgs field takes
up random and independent directions in its internal space in regions of
size $\sim \xi $. Thus, $D_\mu \widehat{\phi }^a\neq 0$. The
correlation length is certainly less than the horizon length at $T_{\textrm{c}}
$, and presumably much less. In reference \cite{Vac91} 
it is argued that these non-zero
covariant derivatives give rise to a non-zero field ${\cal F}_{\mu \nu }^{%
\textrm{em}}$. 

Suppose, therefore, we are 
interested in the magnetic field
component averaged over a line segment $L=n\xi$.  Vachaspati argues
that, since the scalar field is uncorrelated on scales greater than
$\xi$, its gradient executes a random walk as we move along the line, and
so the average of $D_i\widehat{\phi}^a$ over $n$ adjacent lattice points
should scale as $1/\sqrt{n}$.  He then concludes that, since the magnetic
field is proportional to the product of two covariant derivatives of the
Higgs field, it will scale as $1/n$.  This latter conclusion, however, 
overlooks the difference
between ($D_j\Phi ^{\dagger })_{\textrm{rms}}(D_k\Phi )_{\textrm{rms}}$ and $%
(D_j\Phi ^{\dagger }D_k\Phi )_{\textrm{rms}}$.  Enqvist and Olesen noticed
this point, and produced a corrected estimate for the averaged field, $B
\sim B_{\xi}/\sqrt n$, where $B_\xi $ is the initial field strength on 
the scale $\xi_{\rm i}$.  The field strength is order $m_W^2$ at the
electroweak phase transition, and it is correlated on the scale $\xi_{\rm i}
\sim 1/m_W$.  The scale corresponding to the initial coherence length  
at the current epoch is $\xi_{\rm i}/a_{\rm ew}$, with $a_{\rm ew} \sim
3\cdot 10^{-15}$, while the field strength will have decreased in
proportion to $a_{\rm ew}^2$. Thus, with $L\sim 1$ Mpc, $n \sim
10^{25}$, and so the line-averaged field today is between $10^{-19}$ and
$10^{-18}$ Gauss, seemingly of the right order of magnitude.  

However,
the significance of this result is open to doubt on two counts.
Firstly, an objection has already been raised by 
Davidson \cite{Dav96}, who pointed out that the physical 
Higgs field is neutral under U(1)$_{\rm em}$, and therefore 
cannot directly generate electromagnetic currents.  One should 
really regard $D_\mu\widehat{\phi}^a$ as the W field, to
which it reduces in the unitary gauge.  Nonetheless, 
electromagnetic fields can still be sourced by currents of
W particles, for
\begin{equation}
\partial^\nu F^{\rm em}_{\mu\nu} = - \sin\theta_{\rm w} 
F^a_{\mu\nu}D^\nu\widehat\phi^a.
\end{equation}
(There is an additional term on the right hand side for 't Hooft's 
definition). Secondly, as Enqvist himself
points out \cite{Enq94}, surface and volume averages of the 
field give much lower  estimates.  The
physics cannot possibly depend on how we average the magnetic field when
measuring it: thus we should examine what the different averages mean.

\section{Averaging Magnetic Fields}

For cosmological applications, we are interested in the value of the
magnetic field on scales of order 1 Mpc, in some suitably averaged sense.
It is not entirely obvious what one
means by a magnetic field on a scale $L$: does one average over a line, a
surface or a volume? Enqvist and Olesen \cite{EnqOle93} use the line average 
\begin{equation}
B_{(1)}=\frac 1L\int_C{\bf B\cdot dx,}\label{eB1} 
\end{equation}
where $C$ is a curve, which we may take to be a straight line, of 
length $L$. (Vachaspati \cite{Vac91} also implicitly used this average, although in an
incorrect way, as we have explained.) It is also possible to consider an 
average flux
\begin{equation}
B_{(2)}=\frac 1{L^2}\int_S{\bf B\cdot dS.} 
\end{equation}
Lastly, there is a volume averaged field, which is a vector,%
\begin{equation}
{\bf B}_{(3)}=\frac 1{L^3}\int_V{\bf B\ }d^3x. 
\end{equation}
We shall show that only the third of these has any information about the
underlying magnetic field: the line and surface averages always deliver the
same behaviour with scale $L$, regardless of how the field actually behaves.

We resolve the field into its Fourier components:%
$$
{\bf B(x)=}\sum_{{\bf k}}{\bf B}_{{\bf k}}e^{i{\bf k\cdot x}}. 
$$
In a very large region $\Omega $, the summation goes over to the integral $%
\Omega \int d^3k/8\pi^3.$ We then postulate that the field is statistically
random and isotropic, with a correlation length $\xi $. We can therefore
suppose that the average power spectrum goes as 
\begin{equation}
P_{\rm B} \equiv \langle |{\bf B}_{{\bf k}}|^2\rangle 
\propto 
\frac{\xi ^{3}}{\Omega} B_{\xi}^2 (k\xi )^{2p},\qquad (k\xi
\ll 1), \label{ePS} 
\end{equation}
where $k=|{\bf k|.}$ If $p\leq -3/2$, there must be a lower cut-off on $%
k$, otherwise the energy density of the magnetic field diverges. The 
correlation length $\xi$ provides the upper cut-off.

Firstly, we compute the r.m.s. line average (\ref{eB1}) over the volume $%
\Omega $. We take the curve $C$ to be a line of length $L$ in the $x$
direction, centred at ${\bf r}$. Then 
\begin{equation}
B_{(1)}=\sum_{k_1k_2k_3}W(k_1L)B_1(k)e^{i{\bf k}\cdot {\bf r}}, 
\end{equation}
where $W$ is a dimensionless window function. If we weight the entire length
of the line equally, we have 
\begin{equation}
W(k_1L)=\frac{\sin (k_1L/2)}{k_1L/2}. 
\end{equation}
Assuming that statistical averages are identical to space averages, we find 
\begin{equation}
\left\langle B_{(1)}^2\right\rangle =\sum_{k_1k_2k_3}W^2(k_1L)|B_1(k)|^2. 
\end{equation}
Defining $\kappa^2 = k_2^2+k_3^2$, we have 
\begin{equation}
\left\langle B_{(1)}^2\right\rangle \sim \frac {B_\xi^2}{ \xi^3} \int
dk_1W^2(k_1L)\int \langle P_{\rm B}(\kappa^2+k_1^2) . 
\end{equation}
Provided that $p>-3/2$, this integral is well-defined, and for $L\gg \xi $,
we obtain 
\begin{equation}
\left\langle B_{(1)}^2\right\rangle \sim 
B_{\xi}^2 \frac{\xi }{ L}. 
\end{equation}
Thus {\em any} power spectrum with $p > -3/2 $ 
will give the same result, that the
r.m.s.~line-averaged magnetic field decreases as $L^{-1/2}$ as the scale $L$
increases. 

In any case, the line average is not a very useful tool for describing an
averaged magnetic field, as it reveals nothing about the power spectrum of
the magnetic field. The same is true of the surface average, for similar
manipulations give for the average flux 
\begin{equation}
\left\langle B_{(2)}^2\right\rangle 
\sim B_{\xi}^2 \frac{\xi^2}{L^2}, 
\end{equation}
again with the proviso that $p>-3/2$. It is only the volume-averaged field
contains any useful information in general, for in this case 
\begin{equation}
\left\langle B_{(3)}^2\right\rangle \sim \left. k^3
\langle |{\bf B}_{{\bf k}}|^2\rangle\right|_{k=2\pi/L} \sim
B_{\xi}^2\left( \frac
\xi L\right) ^{(2p+3)}. 
\end{equation}
Hogan \cite{Hog83} made a similar point: to calculate the average flux 
on a scale $L$, one must smear the surface of area $L^2$ on a scale
$L$, otherwise the average merely picks out the irrelevant high
frequency power. 
The dynamics of the magnetic field are expressed in terms
of differential equations for the 
individual Fourier components ${\bf B}_{\bf k}(t)$: the 
``seed fields'' are then clearly 
initial conditions for these Fourier components on the 
appropriate scale. 

The line average {\em is} important when calculating
the rotation measure of distant radio sources, which depends on
$\int n_e {\bf B}\cdot d{\bf x}$ along the line of sight, where 
$n_e$ is the free electron density 
\cite{MagF}.  From this can be derived upper bounds on 
cosmological magnetic fields \cite{Kol97}.

What values of $p$ can we expect? In a thermal electroweak phase transition,
as envisaged by Vachaspati, there are several arguments which all 
give $p=0$. 
His original idea was that the mixing between the SU(2) gauge fields and
the hypercharge field allowed electric currents, and hence magnetic fields,
to be generated by gradients of the Higgs field. The Higgs field is
correlated on some scale $\xi _{{\rm H}}$, but not above, and so therefore
is the magnetic field. Averaging ${\bf B}$ over a volume $L^3$, with $L\gg
\xi $, one finds that the individual correlated regions of field add
randomly, and therefore $\int_{L^3}d^3x{\bf B}$ grows only as $(L/\xi _{{\rm %
H}})^{3/2}$. Hence $\left\langle B_{(3)}^2\right\rangle \sim L^{-3}$, which
corresponds to $p=0$. One can also suppose, as did Enqvist and Olesen, that
the magnetic field has its own correlation length,
and therefore the total field averaged over a set of correlation
volumes adds as a random walk. This results in the same answer, $p=0$.
Lastly, $p=0$ is precisely what one obtains from an electromagnetic field in
thermal equilibrium. To see this, let us consider the energy density in
photons up to a frequency $\omega \sim L^{-1}$, which is 
\begin{equation}
\rho _\gamma (L)\sim \int_0^{L^{-1}}d\omega \omega ^3\frac 1{e^{\omega
/T}-1}. 
\end{equation}
For $LT\gg 1$, we find 
\begin{equation}
\rho _\gamma (L)\sim TL^{-3}. 
\end{equation}
However, as the energy is equally distributed between the electric and
magnetic fields, we have also 
\begin{equation}
\rho _\gamma (L)\sim \sum_{{\bf k}}^{L^{-1}}|{\bf B}_{{\bf k}}|^2. 
\end{equation}
Thus a volume-averaged magnetic field in thermal equilibrium has a power
spectrum (\ref{ePS}) with $p=0$. 

Lastly in this section, we note that a scale invariant spectrum, as
might be generated in an almost-de Sitter inflationary model
\cite{Rat92}, corresponds to $p=-3/2$, for which we must also specify a
long-wavelength cut-off $\Lambda$. In this case, all the averages have
the same behaviour with L,
\begin{equation}
\langle B^2_{(a)}\rangle \sim B_\xi^2 \ln(\Lambda/L).
\end{equation}

\section{Causal generation of magnetic fields in the radiation era}
The fact that fields in equilibrium have the same power 
spectrum as that obtained by Vachaspati's mechanism leads one 
to ask how one might distinguish a seed field field generated 
at the phase transition from the background thermal 
fluctuations.  It is natural to suppose that at the phase
transition a linear combination of the fluctuations in the W
and Y fields will emerge as thermal fluctuations in the 
electromagnetic field, with a scale $(eT)^{-1}$. 
We should therefore be careful to 
distinguish them from any putative seed field. A seed 
field should have a scale much greater than the thermal 
fluctuation scale, otherwise it would be absurd to talk of
it being ``frozen in'' \cite{Sacha}.  Its amplitude in each Fourier
mode should also be greater.  A transition 
which remains in thermal equilibrium cannot generate anything
other than thermal fluctuations, which are cosmologically
interesting only in that they evolve into the cosmic 
microwave background.  We must therefore invoke a departure 
from thermal equilibrium by, for example, a first order 
phase transition. This can certainly generate a scale quite 
different than $(eT)^{-1}$ via the average bubble size when 
the low temperature phase percolates.  It is not clear,
however, how much of the energy of the bubbles can be transferred 
into the magnetic field.

Henceforth we free ourselves from any particular mechanism and
try to see if there are bounds on primordial fields from processes
operating in the radiation era from general principles such as energy
conservation. 
Let us suppose that some process at time $t_{\rm i}$ 
creates a magnetic field 
of strength $B_{\rm i}$ on a scale $\xi_{\rm i}$.   Causality 
demands that $\xi_{\rm i} < t_{\rm i}$ and that the power 
spectrum goes as $k^n$, with $n\ge 0$, at small $k$, 
and energy conservation that 
$B^2_{\rm i} < \rho(t_{\rm i})$, the total energy density.  
If the field were completely frozen in, the power spectrum would remain
of the form (\ref{ePS}), with the coherence scale of the field 
$\xi$ fixed in comoving coordinates. The 
energy density (or equivalently the mean square fluctuation) 
on a scale $L$ greater than $\xi=a(t)\xi_{\rm i}/a_{\rm i}$ is given by
\begin{equation}
B^2_{\rm fr}(t,L) \simeq B_{\rm i}^2 \left(\frac{a_{\rm i}}{a(t)}\right)^4 
\left(\frac{a(t)\xi_{\rm i}}
{a_{\rm i} L}\right)^{3+n}.
\end{equation}
This is of course an idealisation: nonetheless $B_{\rm fr}(t,L)$ is a 
useful reference value.

Consider the ratio 
\begin{equation}
r(L)=
B^2(t_0,L)/8\pi\rho_\gamma(t_0),
\end{equation}
where $\rho_\gamma$ is the photon energy density.  From the requirement 
that the magnetic energy density not exceed the total,
\begin{equation}
r_{\rm fr} < \frac{1}{g_*(t_{\rm i})}\left(\frac{\xi_{\rm i}}
{a_{\rm i} L}\right)^{3+n}
\end{equation}
where $g_*(t)$ is the effective number of relativistic degrees of freedom 
at time $t$.  
Thus
\begin{equation}
r_{\rm fr}(L) < \frac{1}{g_*(t_{\rm i})}\left(\frac{T_{\rm eq}}{T_{\rm 
i}}\right)^{3+n} 
\left(\frac{\lambda_{\rm eq}}{L}\right)^{3+n},
\label{eBound1}
\end{equation}
where $\lambda_{\rm eq}$ is the comoving horizon scale at 
$t_{\rm eq}$, the time of equal matter and radiation densities. 
With $H_0 = 50\; {\rm km} \, {\rm s}^{-1} \, {\rm Mpc}^{-1}$, 
$T_{\rm eq} \simeq 1\;{\rm eV}$ and $\lambda_{\rm eq} \simeq 
50\;{\rm Mpc}$.

For the QCD phase transition, $T_{\rm i} \sim 10^8$ eV, and for the 
electroweak transition $T_{\rm i} \sim 10^{11}$ eV. Thus, for 
causal processes happening at these transitions we obtain the 
upper bounds on frozen-in fields 
\begin{equation}
r_{\rm fr}(0.5 \;{\rm Mpc}) < 
\begin{array}{ll}
10^{-19 -6n} & \textrm{(QCD transition)}, \\
10^{-29 - 9n} & \textrm{(Electroweak transition)},
\label{eBound2}
\end{array}
\end{equation}
where we have taken $g_*$ to be 10 and 100 respectively.  Given that 
$r=1$ corresponds to a field strength of $3\cdot 10^{-6}$ Gauss, 
the electroweak transition can at best manage about $10^{-20}$ 
Gauss on the 0.5 Mpc scale.
Thus it can be seen that it is energetically possible 
that some causal process operating at the time of the electroweak 
phase transition could produce a frozen-in seed 
field.  However, nearly the energy density 
of the Universe would have to be converted into a field
 coherent at the horizon scale, which seems 
very hard to arrange. It is perhaps more realistic to suppose that 
the equipartition energy can be transferred to a field coherent on 
$1/10$ the horizon scale, from which we obtain an upper
bound of $10^{-25}$ Gauss.

Magnetic fields are of course not completely 
frozen into the plasma. The
field can drag matter around with it, and field lines can 
pass through each other by reconnection.  The result is a
power-law increase in the comoving scale of the field \cite{Hog83}. 
However, any process which merely increases the scale $\xi$ 
will actually weaken the field on larger scales: in order
to generate stronger fields one must transfer power 
from small to large scales.
By causality, this process cannot operate on 
scales greater than the horizon size $\sim t$, so the power transfer,
if it happens, must take power from below the coherence scale $\xi$ and 
inject into scales between $\xi$ and $t$. 

A set of model MHD equations was shown to have just such an 
``inverse cascade'' 
by Brandenburg et al.\ \cite{BraEnqOle96}.  However, in
the absence of solutions to the full MHD equations one cannot
be fully confident that a primordial field would also show
such behaviour.

\section{Conclusions}
This paper has focused on the generation of fields at the electroweak 
phase transition, but we have arrived at some more general results. 
Firstly, we have tackled the confusion surrounding the concept of
the average magnetic field on a scale $L$, showing that the 
quantity of dynamical interest is the \textit{volume} average of the 
field.  Secondly, we have re-examined Vachaspati's 
proposal for generating fields at the electroweak phase transition. 
It was shown that the argument is based on a particular 
definition 
of the electromagnetic field in the Standard Model, 
which is not unique. In any case, one must be careful 
to distinguish any 
random magnetic seed field from a thermal field. 
The remnant of the thermal radiation at
the electroweak transition is around us in the form of the Cosmic
Microwave Background: Thus a continuous electroweak phase 
transition, where there is essentially no departure from thermal equilibrium,
produces negligible fields on galactic scales. 
Thirdly, we saw that on grounds of causality and energy conservation 
alone, the idea of generating a seed field at or 
before the electroweak transition is quite unlikely, unless there is 
some extra physics in the form of an inverse cascade in the 
turbulent magnetic field, which transfers energy from small-scale 
fluctuations in the field to large ones.  The QCD transition is 
not so strongly constrained, and mechanisms exist based
on charge separation in advanced phase boundaries \cite{SigOliJed97}.

\section*{Acknowledgements}
MH thanks Sacha Davidson, Kandu Subramanian and Tanmay Vachaspati 
for helpful conversations, and Ola T\"ornqvist for 
communicating \cite{Tor97} before publication, which covers
some of the same ground in its discussion of the electromagnetic field
in the electroweak theory.   MH
is supported by the Particle Physics and Astronomy Research Council of
the UK, under an Advanced Fellowship B/93/AF/1642 and grant GR/K55967, and by 
the European Commission under the Human Capital and Mobility
programme, contract no.\ CHRX-CT94-0423.  AEE wishes to thank Ed Copeland 
and the members of the Physics and Astronomy Subject Group at Sussex 
University for their kind hospitality during the commencement 
of this work.

%\end{thebibliography}
\end{document}